\documentclass[twocolumn]{aastex63}

\usepackage{graphicx}
\usepackage{color}
\usepackage{amsmath}  
\usepackage{verbatim}     
\usepackage{hyperref}
\usepackage[noabbrev,capitalise]{cleveref}  
\usepackage{natbib}
\usepackage[mathscr]{euscript} 
\usepackage{xspace}

\makeatletter
\usepackage{etoolbox}
\patchcmd\H@refstepcounter{\protected@edef}{\protected@xdef}{}{}
\makeatother

\newcommand{\sn}{SN~Ia\xspace}
\newcommand{\sne}{SNe~Ia\xspace}

\newcommand{\h}{\ensuremath{\text{H}_0}\xspace}
\newcommand{\kms}{\ensuremath{~\text{km s}^{-1}}\xspace}

\newcommand{\hr}[1][]{Hubble-Lema\^itre residual{#1}\xspace}

\newcommand{\su}{\ensuremath{\sigma_{\mathrm{unexplained}}}\xspace}
\newcommand{\unity}{UNITY\xspace}

\newcommand{\un}[1]{~\text{#1}\xspace}  

\begin{document}

\title{Evidence for Cosmic Acceleration is Robust to Observed Correlations Between Type Ia Supernova Luminosity and Stellar Age}
\shorttitle{Evidence for Cosmic Acceleration is Robust}

\author[0000-0002-1873-8973]{B. M. Rose}
\affiliation{Space Telescope Science Institute, 3700 San Martin Drive, Baltimore, MD 21218, USA}
\author[0000-0001-5402-4647]{D. Rubin}
\affiliation{Physics and Astronomy Department, University of Hawaii, Honolulu, HI 96822, USA}
\affiliation{Physics Division, E.O. Lawrence Berkeley National Laboratory, 1 Cyclotron Road, Berkeley, CA, 94720, USA}
\author[0000-0001-7101-9831]{A. Cikota}
\affiliation{Physics Division, E.O. Lawrence Berkeley National Laboratory, 1 Cyclotron Road, Berkeley, CA, 94720, USA}
\author[0000-0003-2823-360X]{S. E. Deustua}
\affiliation{Space Telescope Science Institute, 3700 San Martin Drive, Baltimore, MD 21218, USA}
\author[0000-0003-1861-0870]{S. Dixon}
\affiliation{Physics Division, E.O. Lawrence Berkeley National Laboratory, 1 Cyclotron Road, Berkeley, CA, 94720, USA}
\affiliation{Department of Physics, University of California Berkeley, 366 LeConte Hall MC 7300, Berkeley, CA, 94720, USA}
\author[0000-0002-6652-9279]{A. Fruchter}
\affiliation{Space Telescope Science Institute, 3700 San Martin Drive, Baltimore, MD 21218, USA}
\author[0000-0002-6230-0151]{D. O. Jones}
\affiliation{Department of Astronomy and Astrophysics, University of California, Santa Cruz, CA 92064, USA}
\author{A. G. Riess}
\affiliation{Space Telescope Science Institute, 3700 San Martin Drive, Baltimore, MD 21218, USA}
\affiliation{Department of Physics and Astronomy, Johns Hopkins University, Baltimore, MD 21218, USA}
\author{D. M. Scolnic}
\affiliation{Department of Physics, Duke University, 120 Science Drive, Durham, NC, 27708, USA}
\correspondingauthor{B. M. Rose}
\email{brose@stsci.edu}
\shortauthors{Rose, et al.}


\received{February 27, 2020}
\revised{May 12, 2020}
\accepted{May 15, 2020}
\submitjournal{The Astrophysical Journal Letters}

\begin{abstract}
Type Ia Supernovae (SNe~Ia) are powerful \replaced{standardized}{standardizable} candles for constraining \deleted{the }cosmological model\added{s} and provided the first evidence of \replaced{accelerated expansion.}{the accelerated expansion of the universe.} Their precision derives from empirical correlations\added{,} now measured from $>1000$ SNe~Ia\added{,} between their luminosities, light-curve shapes, colors and most recently \deleted{a modest relationship }with the \added{stellar }mass of their host galaxy. As mass correlates with other \replaced{host}{galaxy} properties, \replaced{these}{alternative parameters} have been investigated to improve SN~Ia standardization though none have been shown to significantly alter the determination of cosmological parameters. We re-examine a recent claim, based on 34 SN~Ia in nearby passive host galaxies, of a 0.05 mag/Gyr dependence of standardized SN~Ia luminosity on host age which if \replaced{extrapolate}{extrapolated} to higher redshifts, \replaced{might accrue to}{would be a bias up to} 0.25 mag, challenging the inference of dark energy. We reanalyze this sample of hosts using both the original method and a Bayesian hierarchical model and find after a fuller accounting of the \replaced{errors}{uncertainties} the significance \replaced{for}{of} a dependence on age to be $\leq2\sigma$ and $\sim1\sigma$ after \added{the} removal of a single poorly-\replaced{measured}{sampled} SN~Ia. To test the claim that a trend seen in old stellar populations can be applied to younger ages, we extend our analysis to a larger sample which includes young hosts. We find the residual dependence of host age (after all standardization typically employed for cosmological measurements) to be \replaced{$0.0011\pm0.0018$ mag/Gyr ($0.6\sigma$)}{consistent with zero} for 254~SNe~Ia from the Pantheon sample, \deleted{consistent with no trend and strongly }ruling out the large but low significance trend \replaced{claimed from the}{seen in} passive hosts.
\end{abstract}

\keywords{dark energy, distance scale, supernovae: general, supernovae: individual (SN2003ic)}

\section{Introduction}\label{intro}

\replaced{Type Ia supernovae (\sne), the thermonuclear explosion of carbon-oxygen white dwarfs, have been used as cosmic distance indicators for over 30 years. Their observed variability can be empirically corrected (Pskovskii 1977; Phillips 1993; Hamuy et al. 1996; Riess et al. 2016; Perlmutter et al. 1997), allowing for precise luminosity distances resulting in the measurements of the energy density of matter and dark energy, driving the expansion history of the Universe (Garnavich et al. 1998a; Riess et al. 1998; Garnavich et al. 1998b; Perlmutter et al. 1999). Since then, there have been improvements in standardization methods (Jha et al. 2007; Guy et al. 2010; Burns et al. 2011; Mosher et al. 2014) and resulting cosmological measurements (Suzuki et al. 2012; Betoule et al. 2014; Riess et al. 2018; Scolnic et al. 2018; DES Collaboration et al. 2019; Jones et al. 2019; Freedman et al. 2019).}{Type Ia supernovae (\sne), the thermonuclear explosion of carbon-oxygen white dwarfs, are precise cosmic distance indicators \citep{Kowal1968,Pskovskii1969}. Their observed variation in brightness can be empirically corrected \citep{Rust1974,Pskovskii1977,Phillips1993,Hamuy1996d,Riess1996,Perlmutter1997}. This allows their luminosity distances to be used to measure the expansion history of the universe and led to the discovery of cosmic acceleration caused by an unknown force, dark energy \citep{Riess1998,Garnavich1998b,Perlmutter1999}. Since then,  standardization methods have improved \citep{Jha2007,Guy2010,Burns2011,Mosher2014} as have the resulting cosmological measurements \citep{Suzuki2012,Betoule2014,Riess2018b,Scolnic2018,DESCollaboration2019,Jones2019,Freedman2019}.}

As the number of \sne at cosmological distances now exceed 1000, the selection criteria have become more stringent.  Of all the \sne \deleted{that are }observed, roughly 75\% \citep{Scolnic2018} are  used for cosmology. Cosmologically useful \sne are required to have sufficient data:  adequate sampling \replaced{around the peak of}{in order to constrain} the light curve and \deleted{of} the decline rate.  They must also pass 
``quality cuts''  i.e. their parameters should be nearest the centers of the population distributions and thus can be \deleted{precisely }standardized through \deleted{the }empirical correlations. Even after standardization, outliers exist, resulting in outlier rejection tools \citep{Kunz2007,Rubin2015} and even the classification of a new class of transients \citep{Foley2013}. 

The accuracy of \sn cosmological measurements require\deleted{s} the absence of a redshift dependence of the standardized luminosity, which we refer to as luminosity evolution. 
\added{The variation in peak luminosity of \sne may be due to unknown properties of the progenitors.
These could have three effects that concern cosmological measurements. First, these variations in progenitor properties can affect the population demographics. This results in a type of bias discussed in \citet{Scolnic2016}. In addition, many progenitor properties that affect the peak luminosity are already corrected for by the empirical standardization process. Ultimately, luminosity evolution comes from a change in the progenitor system and peak luminosity that is not accounted for in our \sn models.}
\deleted{This is not the same as a change in the mean of the properties used to standardize \sne which may result from sample selection and is commonly referred to as a change in demographics. }

\replaced{Luminosity}{As a proxy for a change in redshift or cosmic time, luminosity} evolution \replaced{may}{can} be constrained locally ($\lesssim 400 \un{Mpc}$) by measuring differences in standardized \sn luminosity between galaxy types. \deleted{that predominately form their progenitor stars recently versus in the distant past as a proxy for a change in redshift or cosmic time. }Over the last decade large samples with strict quality control have revealed correlations between host galaxy properties and standardized peak luminosity \replaced{of modest significance and level}{at a modest level} \citep[e.g.][]{Gallagher2008,Kelly2010,Sullivan2010,Lampeitl2010,Gupta2011,Rigault2013,Jones2015,Moreno-Raya2016a,Uddin2017a,Kim2018,Rigault2018,Jones2018,Rose2019}. Each of these measurements agree in \added{the} direction of the host galaxy effect\replaced{, and}{;} it is clear that these \replaced{are not all}{do not agree} by chance.
Since the average galaxy changes with redshift and sample selection, it has become necessary to include such correlations in the standardization process to limit biases to the $1 \%$ level in distance \added{\citep{Rigault2013}}. \replaced{The first recognized and most commonly used host property for such standardization has been host mass (Kelly et al. 2010; Sullivan et al. 2010; Lampeitl et al. 2010), where there has been seen a $\sim 0.06 \un{mag}$ change in \hr{} across a division in host stellar mass.}{The first recognized and most commonly used host property for such standardization is stellar mass \citep{Kelly2010,Sullivan2010,Lampeitl2010}. This standardization is referred to as the ``mass step'' because of the $\sim 0.06 \un{mag}$ change in average \hr{} at $\sim 10^{10} \un{M}_{\odot}$.} \hr{s} are the difference between the measured luminosity distance and the expected distance from the best-fit cosmology.

\deleted{There are additional tests for luminosity evolution, beyond correlations with galaxy properties. For instance, one can determine if general light curve correlations and \sn spectra behave similar locally and at a redshift of one 
(Scolnic \& Kessler 2016; Foley et al.2008, respectively. 
Like all evidence for luminosity evolution, these changes would be beyond the expected changes in population demographics.}

\defcitealias{Kang2020}{K20}

\citet{Kang2020} (hereafter \citetalias{Kang2020}) claim to \replaced{find}{have found} a correlation between the ages of 34 early-type host galaxies --- derived from spectral features --- and \replaced{their \sn distances}{\sn peak luminosity}. If extrapolated to younger ages and higher redshifts\added{, by} convolving look-back time and \sn progenitor models, \added{this correlation} could cause a \deleted{significant} redshift dependent luminosity evolution, $\Delta \mathrm{mag}/\Delta z > 0.2 \un{mag}$. The original discovery of \replaced{dark energy from}{accelerating cosmic expansion using} \sne \citep{Riess1998,Perlmutter1999} ruled out such a large evolution in standardized luminosity by demonstrating consistency between \sn in early-type hosts and those in young, star-forming hosts.  \deleted{\citetalias{Kang2020} claim a much larger trend between galaxy age and \sne standardized luminosity than seen between host type and \sn standardized luminosity, that along with other assumptions, this would bring into question the evidence for dark energy. However, we have serious concerns about the accuracy of this analysis, and here we show when these are addressed the strong evidence from \sne for dark energy is unchanged.}

\added{\citetalias{Kang2020}'s use of high signal-to-noise spectra to measure metalicities and ages of the host galaxies is impressive, however we have serious concerns about the cosmological interpretation. \citetalias{Kang2020}'s finding of a correlation does not seem to be robust against different sample selections, or different assumptions about uncertainties. In addition, the application of the mass step correction drastically reduces the observed effect in external data.}
\replaced{This does not mean that we disagree with the motivation of the \citetalias{Kang2020} work -- that}{The motivation of the \citetalias{Kang2020} work is well justified --} correlations between \sn properties and their hosts exist, and \deleted{that }these will need to be better characterized to significantly improve upon present cosmological measurements. However, \added{in this work,} we show \deleted{here }that these correlations are not significantly limiting our current ability to use \sne to measure the cosmological parameters of our universe. 


\section{Data \& Techniques}
\explain{The section ``Re-examining the K20 Analysis'' was renamed to better match the paper's new more concise structure.}

\replaced{Between the two paper series \citet{Kang2016} and \citetalias{Kang2020} observed 51 early-type low redshift \sn host galaxy, obtaining extremely high signal-to-noise galaxy spectra (S/N ~ 175, Kang et al. 2016). The \sn are archival, spanning from 1990 to 2010, and aggregated into a uniform re-analysis in the YONSEI SN catalog \citep{Kim2019}. It is the extremely high quality spectra that makes this data set unique, most \sn host galaxy research uses photometry (e.g. Gupta et al. 2011; Jones et al. 2018; Rose et al. 2019)  but some use lower signal-to-noise integral field unit spectra \citep{Rigault2013,Rigault2018}. The methodology used in \citetalias{Kang2020} are not new nor controversial, for they built on previous research, such as \citet{Gallagher2008}.}{Altogether \citet{Kang2016} and \citetalias{Kang2020} observed 51 early-type, low redshift \sn host galaxies, obtaining high signal-to-noise galactic spectra \citep[S/N $\sim 175$,][]{Kang2016}.  The high quality spectra allow for precise measurements of the \sn host galaxy properties. Most \sn host galaxy studies use photometry \citep[e.g.][]{Gupta2011,Jones2018,Rose2019} though some studies use lower signal-to-noise integral field unit spectra \citep{Rigault2013,Rigault2018}. The \sn analyzed by \citetalias{Kang2020} are archival, taking place from 1990 to 2010, and reanalyzed uniformly in the YONSEI SN catalog \citep{Kim2019}. The age measurement techniques used by \citetalias{Kang2020} are well established \citep{Faber1992,Worthey1994}, and built on previous \sn research, such as \citet{Gallagher2008}.}

\deleted{However, our re-examination of the claims by \citetalias{Kang2020} can be grouped into three major areas: first, the quality of the \sn used, second, the robustness and overall significance of the trend in \hr with stellar population age, and third, the extrapolation of this trend to \sne from younger stellar environments, and hence, redshifts.}

\subsection{\sn Data Quality}\label{sec:data}

\added{\citetalias{Kang2020} observed 51 \sn host galaxies. Via various cuts described in their paper, the fiducial analysis was performed with 34 \sn and their host galaxies. Using several definitions of \sn quality, we find 10 of the final 34 \sne fail at least one quality cut. Using just the Joint Light-Curve Analysis (JLA) cosmology cuts \citep[Table 7]{Betoule2014}, nearly $12\%$ (4 out of 34 objects) of the final sample are not of cosmological quality.}

\begin{figure}
    \centering
    \includegraphics[width=0.95\columnwidth]{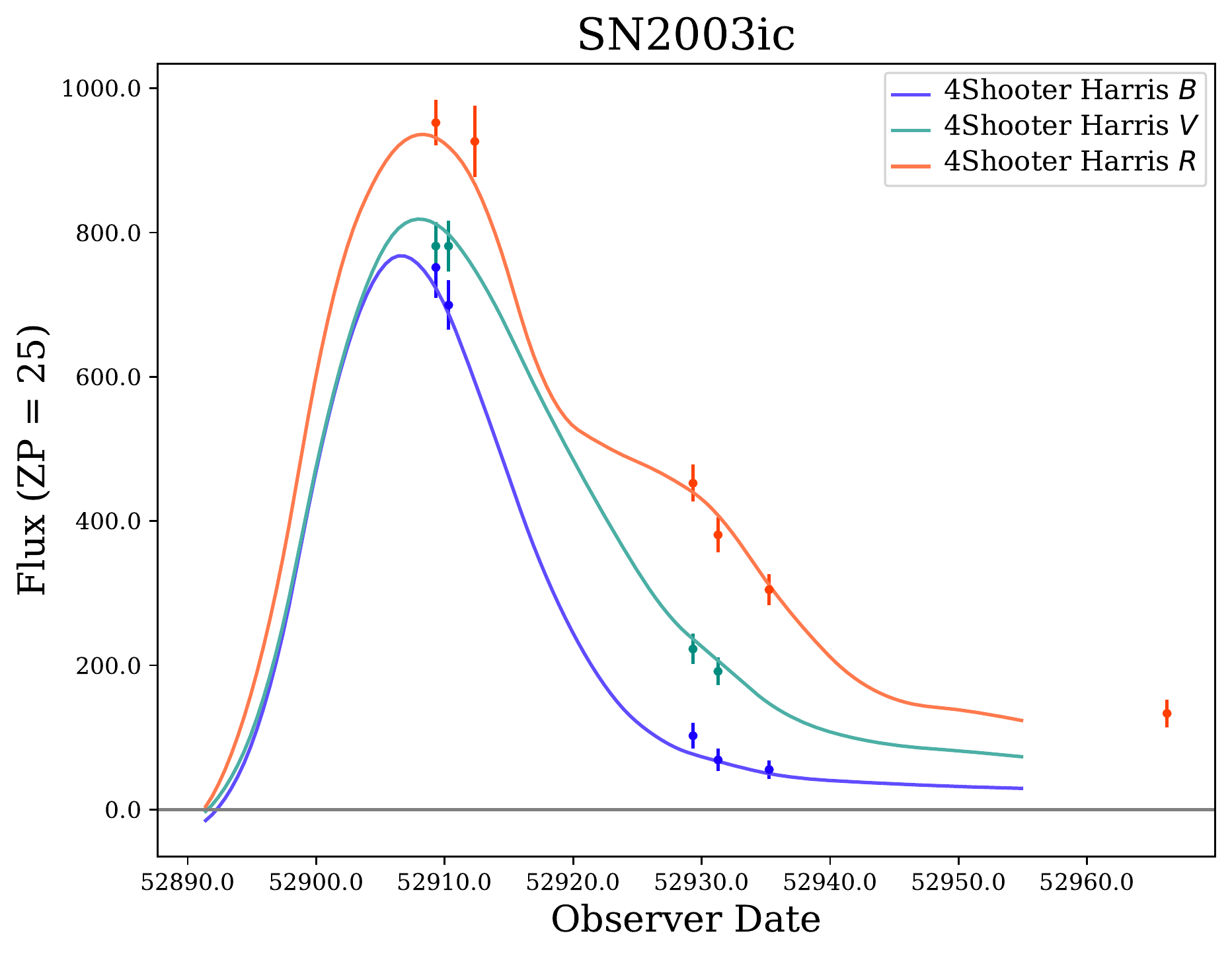}
    \caption{The light curve of SN2003ic. With no pre-maximum data points, it would \replaced{traditionally}{commonly} be removed from cosmological samples due to difficulties constraining the peak luminosity. However this \sn alone takes the correlation seen in \citetalias{Kang2020} from a $1.8\sigma$ to a $2.3\sigma$ significance.
    Original photometry from \citet{Hicken2009}.
    }
    \label{fig:03ic}
\end{figure}

\replaced{The \sn used in \citetalias{Kang2020} are archival, and as such are of varying quality. However, calculating \hr{s} is directly affected by the quality of the \sn data. First we want to bring out that there are several \sn with extremely poorly sampled light-curves and as such would have very uncertain \hr{s}.}{The precision and accuracy of \sn distances depends on the quality of the light-curves of the \sn. There are several \sn in the \citetalias{Kang2020} sample with poorly sampled light-curves, and light-curve fits for these \sne will be problematic.}
\replaced{SN2007ap and SN2008af have no data prior to +5 days post-maximum.
SN2003ch, SN2003ic, SN2003iv, and SN2007cp have less than seven nights observed, with half of these having less than four nights.}{The light curves for SN2007ap and SN2008af have no data prior to five days past maximum. SN2003iv and SN2007cp have fewer than four nights of observations, and SN2003ch and SN2003ic have fewer than seven.} \added{As an example, SN2003ic is shown in \cref{fig:03ic}.}
Finally \added{both} SN1993ac and SN2001ie have \deleted{both} no data prior to \replaced{+5}{five} days post-maximum and \replaced{less}{fewer} than seven nights \replaced{observed}{of observations}.
Of these eight \sn with \replaced{questionable to bad}{poorly sampled} light curves, only SN2003ch and SN2007cp were removed from the final data set \added{of 34 \sn} used in \citetalias{Kang2020}. \deleted{Therefore, the reliability of the \hr measurements of 6 out of 34 \sn can be called into question.}

\replaced{To test assumptions used in cosmological analyses, it is critical that the subset of \sn are reasonable. They do not necessarily need to be a fully representative sample, but they should at least all pass the typical quality cuts.}{To test assumptions used in cosmological analyses, it is not necessary to use a fully representative sub-sample, but they should at least all pass the typical quality cuts.} \replaced{Using the Joint Light-Curve Analysis as an example (JLA, Betoule et al. 2014, Table 7),}{Using JLA as an example,} we see that 4 of the final 34 \sn do not pass quality cuts. \deleted{Several others of the initial \citetalias{Kang2020} \sn sample also fail. }SN2002do and SN2007au \replaced{have}{are best fit with the light-curves shape parameter} $x_1 < -3$. These fast decliners are outside the valid range of the SALT2 model \added{\citep{Guy2007, Guy2010}} requiring alternative \added{standardization} methods \citep[e.g.,][]{Garnavich2004}. In addition, SN2002do, \deleted{SN2004gc, }SN2006kf, and SN2008ia \replaced{are highly extincted by the Milky Way, having}{all have high Milky Way dust extinction, with} $E(B-V)_{MW} \geq 0.15$. The more dimming and reddening from Milky Way dust, the less accurate the \sn peak luminosity can be. For this reason, cosmological analyses typically use \sne that are out of the plane of the Milky Way. \replaced{The Pantheon analysis (Scolnic et al. 2018) performs very similar quality cuts.}{Pantheon and Union perform very similar quality cuts \citep[respectively]{Scolnic2018,Suzuki2012}.}
Other analyses include additional cuts on the phase coverage of the light curves\added{, expressed in terms of rest-frame days from maximum brightness}. \replaced{For example, \citep{Rest2014} requires at least one observation between -10 and +5 rest frame days after maximum brightness, at least one observation between +5 and +20 days after maximum, and at least 5 total observations between -10 and +35 days after maximum.}{For example, \citet{Rest2014} required at least one observation between $-10$ and +5 days, at least one observation between +5 and +20 days, and at least 5 total observations between $-10$ and +35 days.} There are 4 \sne (SN1993ac, SN2001ie, SN2007ap, SN2008af) in the final sample of \citetalias{Kang2020} that fail the first cut, and another \sn (SN2003ic) fails the second.

\begin{deluxetable*}{c|cc|cc|cc}
\tablecolumns{7}
\tablewidth{\textwidth}
\tablecaption{\citetalias{Kang2020} \sn that do not pass typical ``quality cuts''\label{tab:cuts}}
\tablehead{
\colhead{} & \multicolumn{2}{c}{\replaced{LC}{Light Curves} Quality} & \multicolumn{2}{c}{JLA Cuts} & \multicolumn{2}{c}{\citet{Rest2014}} \\
\colhead{} \vspace{-0.5em}& \colhead{$\geq 1$ obs.} & \colhead{Total num obs.} & \colhead{} & \colhead{} & \colhead{$\geq 1$ obs.} & \colhead{$\geq 1$ obs. }\\
\colhead{} & \colhead{$t<+5$ days} & \colhead{$>7$} & \colhead{$|x_1| < -3$} & \colhead{$E(B-V)_{MW} \geq 0.15$} &
\colhead{$-10 < t< +5$ days} & \colhead{$+5 < t< +25$ days}
}
\startdata
SN1993ac\tablenotemark{ac}  & X & X &   &   & X &   \\
SN2001ie\tablenotemark{ac}  & X & X &   &   & X &   \\
SN2002do\tablenotemark{b}   &   &   & X & X &   &   \\
SN2003ic\tablenotemark{ac}  &   & X &   &   &   & X \\
SN2003iv\tablenotemark{a}   &   & X &   &   &   &   \\
SN2006kf\tablenotemark{b}   &   &   &   & X &   &   \\
SN2007ap\tablenotemark{ac}  & X &   &   &   & X &   \\
SN2007au\tablenotemark{b}   &   &   & X &   &   &   \\
SN2008af\tablenotemark{ac}  & X &   &   &   & X &   \\
SN2008ia\tablenotemark{b}    &   &   &   & X &   &   \\
\enddata
\tablenotetext{a}{Has a poorly sampled light curve.}
\tablenotetext{b}{Fails JLA quality cuts, defined in \citet{Betoule2014}.}
\tablenotetext{c}{Fails phase coverage cuts, defined in \citet{Rest2014}.}
\tablecomments{From the final 34 \sn in the \citetalias{Kang2020} sample, 6 \sn have poorly sampled light curves, 4 would not pass the JLA cuts, and 5 would not pass the \citet{Rest2014} phase coverage cuts.
}
\end{deluxetable*}

A summary of which \sn fails what cut can be seen in \cref{tab:cuts}. \deleted{Using several definitions of \sn quality, we find 10 \sne that fail at least one quality cut. Using just the JLA cosmology cuts, nearly $12\%$ (4 out of 34 objects) of the final sample is not of cosmological quality, resulting in unreliable \hr{s}. Our first conclusion is that \citetalias{Kang2020} properly cleans the data for defects in the host galaxies, but does not perform the same treatment to their \sne.}

\subsection{Standardization and Uncertainties}\label{sec:correlation}
\explain{The subsection ``Correlating Between Hubble-Lema\^itre Residual and Age" was renamed to better represent the new organization.}

\deleted{The reduced $\chi^2$ statistic ($\chi^2_{\nu}$) of the fiducial \citetalias{Kang2020} correlation is $\sim$2 indicating significant unaccounted for uncertainty.} For \sn at low redshift (the \citetalias{Kang2020} sample is at $z<0.04$) \replaced{this is typically from}{there are several important uncertainties to consider:} \added{the uncertainty in the} local peculiar motion\replaced{ with an uncertainty $\sigma_v$,}{ ($\sigma_v$)} and the unexplained scatter seen in \sn post standardization (\replaced{$\sigma^{\mathrm{mag}^2}_{\mathrm{unexplained}}$}{$\sigma_{\mathrm{unexplained}}$}).\footnote{\citetalias{Kang2020} uses a common alternative name, intrinsic dispersion ($\sigma_{\mathrm{int}}$).} \replaced{Further, if}{If} one accounts for expected flows using maps of large-scale structure on a \sn-by-\sn basis\deleted{,} as undertaken by \citetalias{Kang2020}, a peculiar velocity uncertainty floor remains due to the unpredictable motions local to each \replaced{\sn}{host galaxy}. Pantheon \citep{Scolnic2018}, \replaced{calculate}{calculated} this to be $\sigma_v=250 \kms$\added{.} \deleted{after estimating velocity corrections. An certainty floor like this significantly increases distance uncertainties at these low redshifts and reduces the significance of correlations.} The total distance uncertainty of a \sn is comprised of many individual uncertainties. A relevant example, based on the Pantheon \replaced{analyse}{analysis} of \citet{Scolnic2018}, is 
\begin{equation}\label{eqn:uncert}
    \sigma_{\mathrm{total}}^2 = \sigma_{N}^2 + \sigma_{z}^2 + \mathbf{\sigma_{v}^2} + \mathbf{\sigma^{^2}_{\mathrm{\textbf{unexplained}}}}
\end{equation}
where $\sigma_{N}^2$ is the photometric error of the \sn distance and
$\sigma_{z}^2$ is the uncertainty from the redshift.
\deleted{The last two terms (in bold) were \replaced{ignored}{misunderstood} in \citetalias{Kang2020}. }\added{In \citetalias{Kang2020}, $\sigma_{\mathrm{unexplained}}$ was misunderstood and $\sigma_{v}$ was absent.}  \deleted{These ignored uncertainties increase the total distance uncertainty by $\sim \sqrt{2}$, would increase the statistical significance of any measured correlation.}

\explain{The following paragraph was moved to this location to better fit within the improved section headers.}
\added{When looking for a trend between \hr{s} and a host galaxy property \added{one} can accidentally ignore cross correlations with the \sn standardization terms \citep{Hamuy1995,Hamuy2000,Smith2020}. Therefore, to further test the observed trend in \citetalias{Kang2020}, we sampled a simple standardization equation in the Bayesian hierarchical model \unity\footnote{\url{https://github.com/rubind/host_unity}} \citep{Rubin2015,Rose2020a}. We used a typical Tripp-like linear standardization \citep{Tripp1998}:
\begin{equation}
    \mu = m_B - \Big(M_B + \alpha x_1 +\beta c + \gamma a\Big)
\end{equation}
where $\mu$, $m_B$, $M_B$ are the distance modulus, apparent and absolute magnitude respectively. The $\alpha$, $\beta$, and $\gamma$ parameters are the linear standardization coefficients corresponding to the SALT2 \citep{Guy2007,Guy2010} light-curve shape ($x_1$) and color ($c$), along with the host galaxy age in gigayears ($a$). The parameters $m_B$, $x_1$, $c$, and $a$ are unique for each \sn, whereas $M_B$, $\alpha$, $\beta$, and $\gamma$ are fit for simultaneously along with any cosmological parameters of interest. \unity also simultaneously fits for the remaining unexplained scatter ($\sigma_{\mathrm{unexplained}}$) allowing for the additional term, $\gamma a$, to explain more of the observed \sn variability but still tracking all uncertainties.}

\section{Re-examining the \sn-Age Correlation}

\subsection{The Impact of SN2003ic}
\explain{This section and subsections were created as apart of the reorganization in order to present our results more concisely.}

\explain{This paragraph and the next were switch relative to the original draft.}
The measured \hr-age trend \citepalias[Figure 13]{Kang2020} visually appears to be \replaced{strongly dependent on}{dominated by} SN2003ic, the \sn with the oldest host. As seen in \cref{fig:03ic} and addressed in \cref{sec:data}, the light curve of SN2003ic is \deleted{very }poorly sampled, including no pre-maximum measurements and only two epochs closely spaced in time to sample the the first 15 days of decline, the most valuable span of time for calibrating the light curve decline rate.
If SN2003ic was removed, the trend shifts from $-0.051 \pm 0.022~\mathrm{mag}/\mathrm{Gyr} ~(2.3\sigma)$ to a less significant $-0.045 \pm 0.024~\mathrm{mag}/\mathrm{Gyr} ~(1.8\sigma)$ using the original \citetalias{Kang2020} data\deleted{ and $1.5 \sigma$ when attempting to get the $\chi^2_{\nu} \sim 1$, as discussed previously}. 
Removing other poorly sampled \sn \replaced{does}{do} not \replaced{effect}{affect} the trend as much as SN2003ic.
A summary of each \hr stellar age correlation discussed in this paper, can be found in \cref{tab:summary}.

\begin{deluxetable*}{lCCc}
\tablecolumns{5}
\tablewidth{0pt}
\tablecaption{Summary of \replaced{the}{discussed} correlations between \hr and stellar age\label{tab:summary}}
\tablehead{
\colhead{Method} & \colhead{Correlation} & \colhead{Significance} & \colhead{Num. \sn}\\
\colhead{} & \colhead{[$\mathrm{mag}/\mathrm{Gyr}$]} & \colhead{}
}
    
\startdata
\citetalias{Kang2020} \added{fiducial analysis} & $-0.051 \pm 0.022$ &  $2.3\sigma$ & 34\\
\citetalias{Kang2020} reproduction & $-0.051 \pm 0.023$ &  $2.3\sigma$ & 34\\
\citetalias{Kang2020} reproduction w/o SN2003ic & -0.045 \pm 0.024 & 1.8\sigma & 33 \\ 
\\
\citetalias{Kang2020} plus 250 km/sec velocity uncertainty & -0.047 \pm 0.022 & 2.1\sigma  & 34\\
above plus 0.10 mag floor on $\sigma_{\mathrm{unexplained}}$ & -0.046 \pm 0.024 & 1.9\sigma & 34\\
above w/o SN2003ic & -0.037 \pm 0.025 & 1.5\sigma & 33\\
\\
UNITY & -0.035 \pm 0.023  & 1.5\sigma & 34\\
UNITY w/o SN2003ic & -0.013 \pm 0.022 & 0.6\sigma & 33\\
\\
Spearman correlation coefficient & \nodata & 2.0\sigma & 34 \\
\\
Pantheon \hr{s} & -0.016 \pm 0.031  & 0.5\sigma & 27\\
Pantheon w/o SN2003ic & +0.008 \pm 0.030 & 0.3\sigma & 26\\
\enddata
\end{deluxetable*}

\subsection{Underestimating Uncertainties}

\added{\citetalias{Kang2020} states that they fit their correlations using \texttt{LINMIX} \citep{Kelly2007a}. This methodology contains an ``intrinsic random scatter,'' counter to the claim that \citetalias{Kang2020} uses no intrinsic scatter. Our reproduction of their work was performed using the \texttt{linmix\_err} package in IDL. We conclude that \su was calculated by \texttt{LINMIX} and was $\sim 0.10 \un{mag}$, as seen in other works. Adding in the peculiar velocity uncertainty, $\sigma_v$, we calculate the significance of the age trend becomes $1.9\sigma$ --- or $1.5 \sigma$ when removing SN2003ic.}

\deleted{If a new correlation were to explain the unexplained scatter, obviating the need to include it in the uncertainty, the result should be a value of $\chi^2_{\nu} \sim 1$ which is not the case for the \citetalias{Kang2020} analysis.
\citetalias{Kang2020} did not include any unexplained scatter in \sne luminosities post-standardization or a peculiar motion error floor.
Their inclusion is important because correctly estimating the significance of any trend is tied directly to accurate accounting of the uncertainty in the data and a realistic error in a trend is only determined if $\chi^2_{\nu}$ $\sim$ 1. Including both $\sigma_v=250 \kms$ velocity uncertainty and an unexplained scatter of $\sim 0.10 \un{mag}$ as used in most cosmological analyses (Suzuki et al. 2012; Betoule et al. 2014; Rubin et al. 2015; Scolnic et al. 2018; DES Collab- oration et al. 2019) brings $\chi^2_{\nu} \sim 1$ for the \citetalias{Kang2020} sample and the significance of the age trend becomes $1.9\sigma$.}

\deleted{In addition, the measured \hr-age trend (K20, Figure  13) visually appears to be strongly dependent on SN2003ic, the \sn with the oldest host. As seen in \cref{fig:03ic} and addressed in \cref{sec:data}, the light curve of SN2003ic is very poorly sampled, including no pre-maximum measurements and only two epochs closely spaced in time to sample the the first 15 days of decline, the most valuable span of time for calibrating the light curve decline rate.
If SN2003ic was removed, the trend shifts from $-0.051 \pm 0.022~\mathrm{mag}/\mathrm{Gyr} ~(2.3\sigma)$ to a less significant $-0.045 \pm 0.024~\mathrm{mag}/\mathrm{Gyr} ~(1.8\sigma)$ using the original \citetalias{Kang2020} data and $1.5 \sigma$ when attempting to get the $\chi^2_{\nu} \sim 1$, as discussed previously. 
Removing other poorly sampled \sn does not effect the trend as much as SN2003ic.
A summary of each \hr stellar age correlation discussed in this paper, can be found in \cref{tab:summary}.}

\explain{We have consolidated our conclusions to be once per subsection.}
\deleted{As both are $< 2\sigma$ our second conclusion is that with realistic errors or the exclusion of a single poorly observed \sn (i.e., as is seen in a basic jack knife test), there is no statistically significant trend with age in the \citetalias{Kang2020} data.}

\explain{This paragraph was moved up in the text to better fit within the improved section headers.}
\deleted{However, looking for a trend between \hr{s} and any host galaxy property \added{one} can easily ignore cross correlations with the \sn standardization terms \citep{Hamuy1995,Hamuy2000,Smith2020}. Therefore, to further test the observed trend in \citetalias{Kang2020}, we sampled a simple standardization equation in the Bayesian hierarchical model \unity\footnote{\url{https://github.com/rubind/host_unity}} \citep{Rubin2015,Rose2020a}. We used a typical Tripp-like linear standardization \citep{Tripp1998}:
\begin{equation}
    \mu = m_B - \Big(M_B + \alpha x_1 +\beta c + \gamma a\Big)
\end{equation}
where $\mu$, $m_B$, $M_B$ are the distance modulus, apparent and absolute magnitude respectively. The $\alpha$, $\beta$, and $\gamma$ parameters are the linear standardization coefficients corresponding to the SALT2 \citep{Guy2010} light-curve shape ($x_1$) and intrinsic color ($c$), as well as the host galaxy age in gigayears ($a$). The parameters $m_B$, $x_1$, $c$, and $a$ are unique for each \sn, whereas $M_B$, $\alpha$, $\beta$, and $\gamma$ are data set variables that are fit simultaneously along with the cosmological parameters of interest. \unity also fits for the remaining unexplained scatter ($\sigma^{\mathrm{mag}}_{\mathrm{unexplained}}$) allowing for the additional term, $\gamma$, to explain more of the observed \sn variability without over estimating the significance of any parameter.}


\explain{The two corner plots were removed to make the paper more concise.}
\deleted{The resulting \unity parameter estimation, using the fiducial sample from \citetalias{Kang2020}, is shown in \cref{fig:UNITY}.} 
\replaced{Using}{When re-analyzing the original data with} \unity, the significance of the \citetalias{Kang2020} trend with age ($\gamma$) is reduced to  $1.5\sigma$.
\replaced{These results from \unity suggest that the significance of any \hr-host galaxy correlations are typically over estimated.}{This suggests that the significance of any \hr-host galaxy correlation is typically over estimated compared to when all parameters are simultaneously fit.} 
As is necessary for an accurate error estimation, we included the non-diagonal covariance terms from the light-curve fitting; \citetalias{Kang2020} only \replaced{report}{reported} diagonal covariance terms. Due to this missing data and some inconsistencies between the values reported in \citetalias{Kang2020} and the original YONSEI SN catalog \deleted{\citep{Kim2019} }(\replaced{e.g.}{i.e.} the $x_1$ value of SN2002G), we used the results from our own light-curve fits.
\deleted{Without SN2003ic, a non-zero $\gamma$ drops to only a $0.6\sigma$ significance.} 
\explain{We have consolidated our conclusions to be once per subsection.}
\deleted{Our third conclusion, like the second, is that the simultaneous consideration of the standardization parameters, appropriate when considering a new parameter, shows that standardizing with host galaxy age is not statistically significant, nor robust to the removal SN2003ic.}

\explain{We removed this paragraph because upon comments from reviewer two and our subsequent investigation we found that this model was not credible.}
\deleted{The high $\chi^2_{\nu}$ values seen in \citetalias{Kang2020} indicates the possible need for additional uncertain in the age measurements.
To test this, we added an additional parameter to \unity ($\sigma^{\mathrm{age}}_{\mathrm{unexplained}}$). This parameter is added in quadrature to each quoted age uncertainty term, resulting in a new total age uncertainty and is modeled off the standard method of adding $\sigma_v$ to the total uncertainty, \cref{eqn:uncert}. If $\sigma^{\mathrm{age}}_{\mathrm{unexplained}}$ is consistent with zero, then the reported age uncertainties should be considered realistic. 
\Cref{fig:ageUncert} shows a highly significant non-zero additional age uncertainty. On average, the uncertainties on the YEPS (Yonsei Evolutionary Population Synthesis, Chung et al. 2013) ages do not fully explain the variance by $1.6 \pm 0.3 \un{Gyr}$. As expected, with proper uncertainties, the significance of $\gamma$ falls to $0.5\sigma$.
Our fourth conclusion is that the uncertainties reported from YEPS are underestimated, on average, by $1.6 \pm 0.3 \un{Gyr}$, and as a result inflating the significance of any non-zero parameter measurement.}

However, we are able to ignore the disputed uncertainties \replaced{($\sigma^{\mathrm{age}}_{\mathrm{unexplained}}$,  $\sigma_{v}^2$, and $ \sigma^{\mathrm{mag}^2}_{\mathrm{unexplained}}$)}{($\sigma_{v}$ and $ \sigma_{\mathrm{unexplained}}$)} and measure a correlation's significance directly from the scatter in the data. This is done via correlation coefficients. The Pearson correlation coefficient is the most common, but assumes both that the trend is linear and that each data set is normally distributed. \replaced{There is no expectation that the age values would be normal, in fact,  Childress et al. (2014) predicates them to be non-normal, and Rose et al. (2019) and others have observationally confirmed that prediction. The Spearman rank-order correlation coefficient does not have these requirements.}{Since the age values have been found to not be normally distributed \citep{Childress2014,Rose2019}, we use the Spearman rank-order correlation which does not have this requirement.} When using the final data set of \citetalias{Kang2020}, the Spearman correlation coefficient is $r_s = -0.35$\added{, a $2.0\sigma$ non-zero result. }\deleted{, slightly higher than the $r_s = -0.23$ correlation seen in \citet{Rose2019}. Uncorrelated variables producing a data set that has a Spearman correlation at least as extreme, is possible at the $2.0\sigma$ level, slightly lower than the $2.1\sigma$ significance seen in the larger \citet{Rose2019} sample. }\added{This result is statistically consistent with the larger data set of \citet{Rose2019}.} Bypassing any question about the accuracy of the uncertainties, this trend appears only marginally significant.

Via several alternative analysis methods --- both accounting for additional known uncertainties and bypassing them --- we have seen the correlation is at most $2\sigma$, but likely less. \deleted{Three of our previously stated conclusions show a reduced certainty and significance of this trend.
}\added{We conclude that there is no statistically significant trend with age in the \citetalias{Kang2020} data.}


\subsection{Extrapolation to Constraints on Cosmological Parameters}
\explain{We renamed the subsection ``Propagating to Cosmology" to better fit the structure.}

Our \replaced{second}{next} set of concerns are based around how \citetalias{Kang2020} extrapolates a correlation with age to a bias in cosmology. As discussed \replaced{in \cref{sec:correlation}}{previously}, the correlations between \hr and host galaxy age is \deleted{highly }dependent on a unique data set of \citetalias{Kang2020} that is not typical of cosmological samples.

\explain{As one of the weakest arguments, we removed this to help make the letter more concise.}
\deleted{First, \citetalias{Kang2020} use a simple argument using the cosmic star formation rate and the \sn delay time distribution to predict the mean progenitor age of a \sn as a function of its redshift (K20,Figure 15). We note that the nature of \sn progenitors remains \deleted{highly} uncertain and the correct such model will depend on whether \sn arise from the merger of two white dwarfs or accretion from a companion star onto a white dwarf. Thus, such a prediction is \deleted{highly} speculative and \citetalias{Kang2020} do not consider any additional uncertainty in their statement of luminosity evolution that the average change in \sn progenitor age is  ``$\sim 5.3 \un{Gyr}$" between $z=0$ and $z=1$. In spite of its precise mathematical derivation, measuring any galaxy physical property is inherently very difficult. Stellar mass is often quoted with an uncertainty of at least $0.3\un{dex}$ (a factor of 2), and it is the simplest \replaced{physical}{derived} property to estimate. When considering stellar ages, a similar limit in the accuracy of the measurements should be assumed. As discussed previously, \deleted{\cref{fig:ageUncert},} it appears that the YEPS ages are under quoted on average by $1.6 \pm 0.3 \un{Gyr}$.}

\replaced{Secondly, modern}{Recent} \sn cosmology analyses \citep{Suzuki2012,Betoule2014,Rubin2015,Scolnic2018,DESCollaboration2019}, all of which have demonstrated strong evidence for \replaced{accelerating expansion and dark energy}{cosmic acceleration}, account for the well-established change in average \hr across a division in host stellar mass.
\replaced{A procedure which reduces the effect from any correlation with age, due to galaxy scaling relationships.}{This procedure reduces the effect of any new correlation with age, due to galaxy scaling relationships. However, the \citetalias{Kang2020} sample uniquely isolates age from stellar mass and morphology.}
\replaced{In addition, many}{Many comological} analyses include a parameter to marginalize over the uncertainty that this change in \hr could be caused by another host galaxy property, such as age. This marginalization would further \replaced{reducing}{reduce} the effect of a trend with age. 
In \citet{Rubin2015}, this marginalization was done with a the redshift dependent mass step\replaced{. The $\delta(\infty)$ parameter did slightly prefer an age-like correlation, drastically reduces the maximum bias of any cosmological parameter due to the correlation reported by \citetalias{Kang2020}.}{, and the resulting best fit cosmology slightly favored an age-like redshift dependence over a pure stellar mass effect. This drastically reduces the maximal bias on cosmological parameters possible from the correlation reported by \citetalias{Kang2020}.}

\begin{deluxetable}{cDCC}
\decimals
\tablewidth{0.95\columnwidth}
\tablecaption{\hr{s} and ages for 254 low redshift Pantheon \sn.\label{tab:data}}
\tablehead{
\colhead{\sn} & \twocolhead{HR} & \colhead{uncertainty} & \colhead{Age}\\
\colhead{} & \twocolhead{[mag]} & \colhead{[mag]} & \colhead{[Gyr]}
}
\startdata
2001ah        & -0.04  & 0.13 & 1.714\\
2001az        & 0.14   & 0.12 & 1.041\\
2001bf        & -0.10  & 0.17 & 1.261\\
2001da        & -0.04  & 0.14 & 1.261\\
2001eh        & -0.01  & 0.16 & 1.924\\
\enddata
\tablecomments{
\hr{s} are from \citet{Jones2018}. Ages (light-weighted) are estimated using ZPEG. We used a fixed 15\% uncertainty in this analysis.
\Cref{tab:data} is published in its entirety online in a machine-readable format. A portion is shown here for guidance regarding its form and content.
}
\end{deluxetable}

To further investigate if \added{a} standard cosmological analysis\replaced{, which}{ that} account\added{s} for both host and selection effects\deleted{,} may mitigate the effect of \citetalias{Kang2020}'s trend on cosmological parameters, we replaced the \hr with those calculated \replaced{by}{during} the Pantheon analysis \citep{Scolnic2018}\replaced{ for the 27 \sn where both were available.}{. There are 27 \sne with both Pantheon \hr{s} and YEPS host galaxy ages.} 
\replaced{The associated \hr{s} and ages are a subset of the full low redshift Parthenon \sn, listed in \cref{tab:data}.}{We present the \hr{s} and ages for the entire low redshift Pantheon sample in \cref{tab:data}.}
Using the standard cosmological correction for the host mass step and \added{a new} observational bias \added{correction framework} \citep[BBC,][]{Kessler2017}, the trend with \hr becomes $-0.016 \pm 0.031~\mathrm{mag}/\mathrm{Gyr}$ or \replaced{a non-zero significance of $0.5\sigma$}{consistent with zero}. 
Without SN2003ic, the trend with the Pantheon's \hr{s} reverses direction ($+0.008 \pm 0.030~\mathrm{mag}/\mathrm{Gyr}$). \deleted{Thus we find no evidence for an age trend from the \citetalias{Kang2020} ages using the distances calculated for cosmological analyses.
}This is also true for the two other age methods used in \citetalias{Kang2020}: going from \citetalias{Kang2020} to Pantheon \hr{s} \replaced{decreases the significance of each correlation to $\sim 1.0\sigma$}{the trend becomes consistent with zero}. \replaced{Our fifth conclusion is}{We conclude} that using \hr{s} that are \replaced{fully standardized, including a host galaxy stellar mass term, the trend is small and insignificant}{standardized with the mass step results in an insignificant trend} and therefore does not propagate to a bias in cosmological estimates.

\subsection{Consistency with Other Data Sets}
\explain{The ``Beyond the K20 sample" section has been changed into a subsection to match the more concise structure.}

\replaced{K20 ultimately apply their trend to \sne in young hosts (the most common hosts) and to \sne at higher redshifts by assuming it can be extrapolated from old \sn to younger stellar populations.}{\citetalias{Kang2020} ultimately applied their trend to cosmological distances by assuming it could be extrapolated to \sne from younger stellar populations.} This interpretation assumes that the physical mechanism is a smoothly varying process rather than discrete \deleted{like multiple }sub-populations \added{as} seen in \citet{Rigault2013} and \citet{Cikota2019}. Indeed, it is quite possible that at all redshifts most \sne are from young progenitors as \sne in early-type hosts galaxies (typically dominated by old stars) make up only a small fraction of cosmological samples \added{\citep{Childress2014}}.

\begin{figure}
    \centering
    \includegraphics[width=0.85\columnwidth]{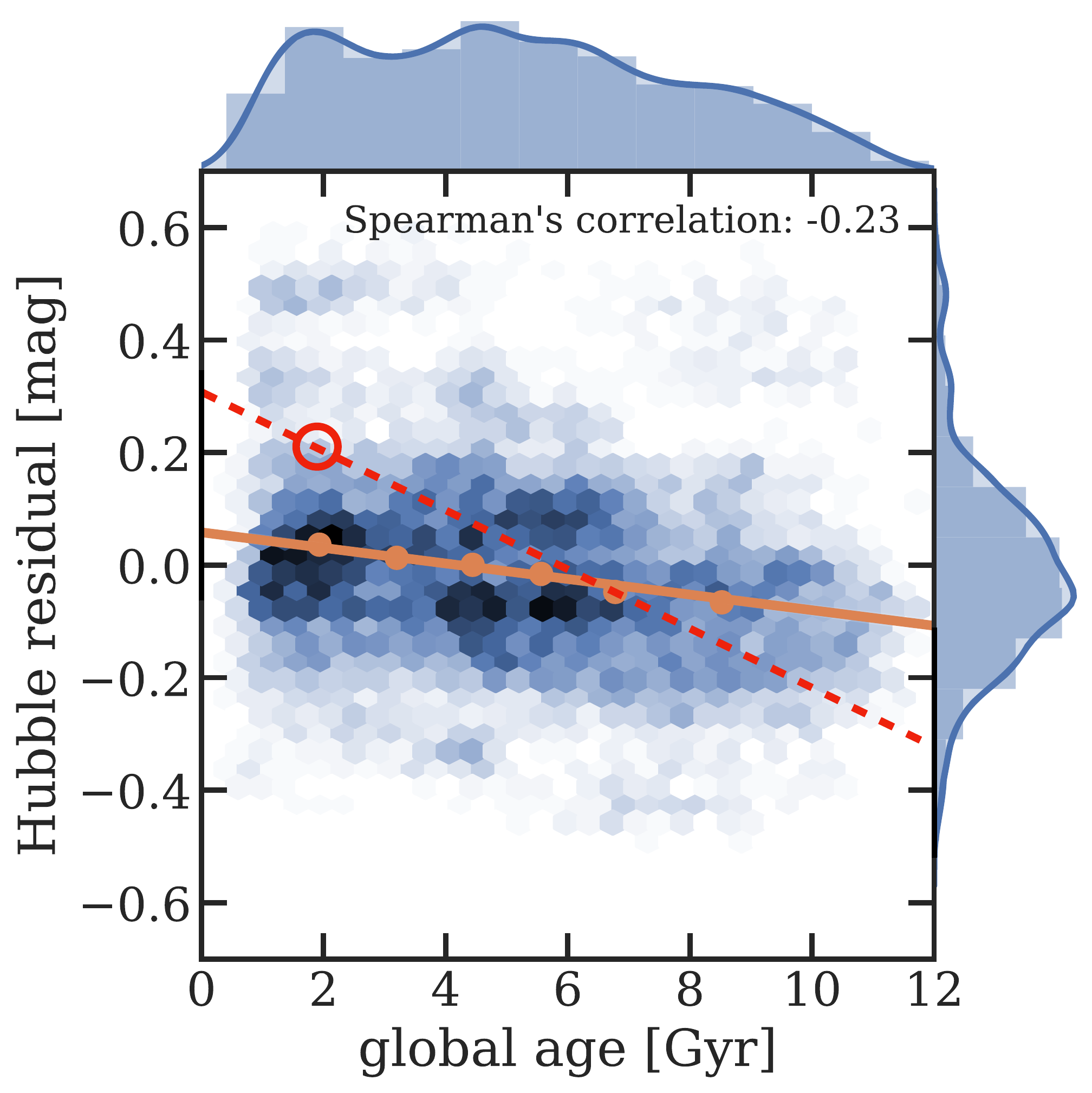}
    \caption{2D density plot (darker colors indicate a higher density) depicting the probability of finding a \sn at a given \hr and average stellar age. \added{The global age is the mass-weighted average age from the galaxy SED. Analysis details can be found in \citet{Rose2019}.} A linear fit to the data (orange line) is shown, along with six evenly filled bins (orange points). The extrapolated trend of \citetalias{Kang2020} is shown as a red dashed line. The predicted average \hr for young hosts is shown as the red circle. This prediction (red circle) is \deleted{highly }inconsistent with the measured average (orange point). The original \added{data and} figure \replaced{is}{are} from \citet{Rose2019}. We note that like \citetalias{Kang2020}, the \hr{s} in \citet{Rose2019} do not include the mass step correction.
    }
    \label{fig:rose2019}
\end{figure}

The interpretation in \citetalias{Kang2020} implies that \sn in young hosts will have an average \hr of $\sim0.25 \un{mag}$. This \deleted{highly }biased average \hr is \deleted{also }ruled out by the analyses of both \citet{Gupta2011} and \citet{Rose2019} who independently looked at data from the Sloan Digital Sky Survey \citep{Sako2008,Campbell2013,Sako2014} using two distinct age estimators. An example of this discrepancy between external data and \citetalias{Kang2020}'s predication can be seen in \cref{fig:rose2019}. Measurements of the \hr{s} for \sne from young host galaxies place the prediction of \citetalias{Kang2020} \deleted{well out }in the tail of the distribution.
\deleted{Our sixth conclusion is that the linear extrapolation to young ages is highly inconsistent with external data.}

The \added{mass-weighted} ages derived from the optical spectral energy distribution (SED) fitting of \citet{Rose2019} are not \replaced{a}{as} precise for any one individual host galaxy as the \citetalias{Kang2020}'s \replaced{YEPS}{YEPS \citep[Yonsei Evolutionary Population Synthesis,][]{Chung2013}} ages derived from spectral features. However, when aggregated, SED based ages are statistically powerful\added{, until they reach the systematic limits of the stellar population models}. \replaced{Just like how photometric redshifts are more uncertain for any one object, in aggregate they can be a powerful tool, so are SED based ages.}{Just like photometric redshifts, SED based ages can catastrophically fail for any one object, but in population studies they are a powerful tool.}

\begin{figure}
    \centering
    \includegraphics[width=0.95\columnwidth]{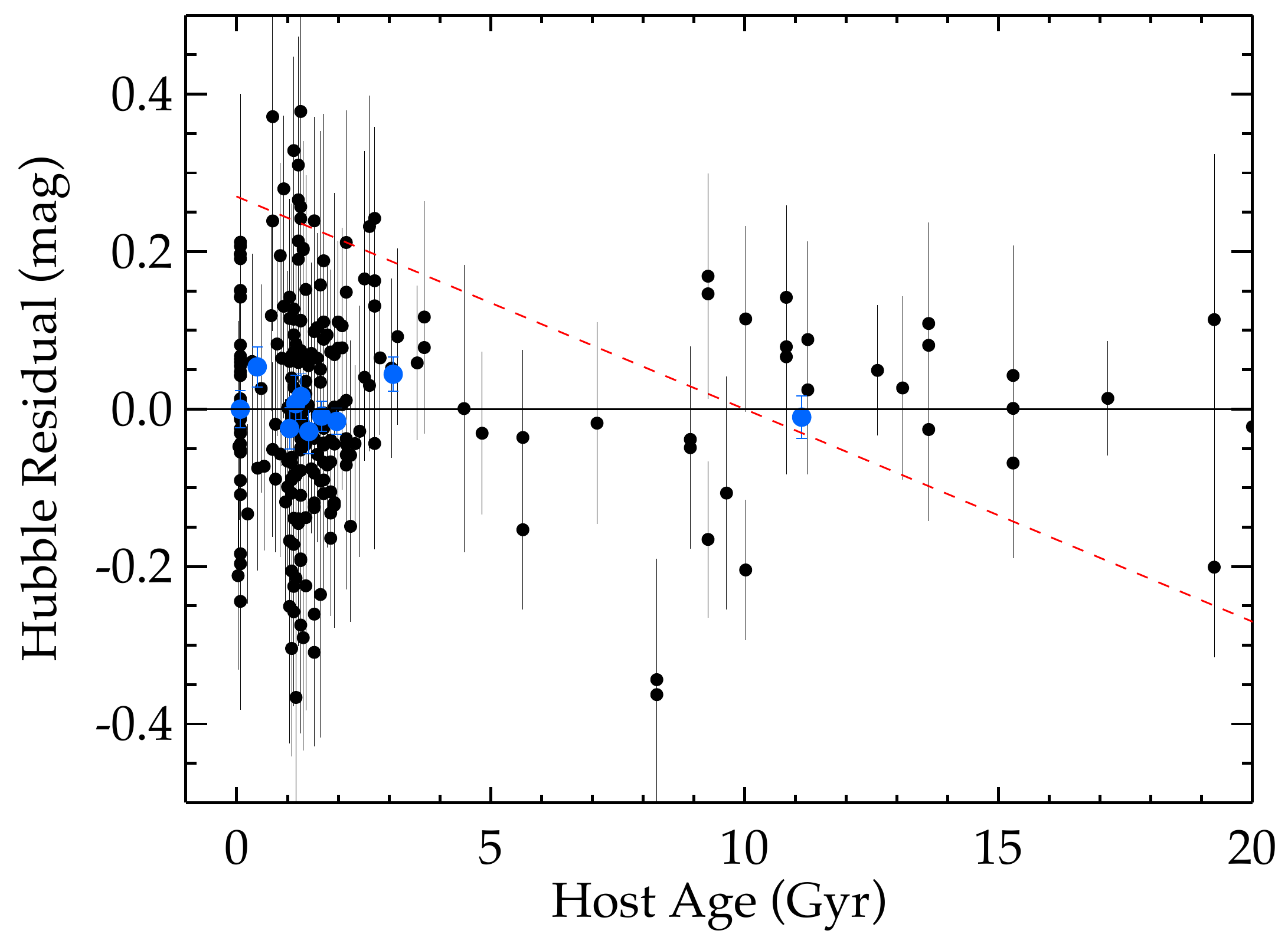}
    \caption{
    The relationship between the cosmological Pantheon sample's \hr{s} and host galaxy age (light-weighted) for the low redshift \sne (black) of \citet{Jones2018}. Blue points are bins of 25 \sn. \deleted{The measured correlation is $0.0011 \pm 0.0018~\mathrm{mag}/\mathrm{Gyr}$ (red line, dashed and dotted red are $\pm 1 \sigma$). }Light-weighted ages are \deleted{typically }biased by bright young stars, reducing the range of observed ages and increasing the measured slope. The \replaced{green}{red} dashed line is the trend seen in \citetalias{Kang2020}. The mass step that is applied to the Pantheon \hr{s} would not drastically shift the trend from \citetalias{Kang2020} because it was derived using only massive early-type galaxies.
    However, one can easily imagine concluding a low-significance trend if one only used hosts with ages $>2.5 \un{Gyr}$ (the last two bins), as expected in a passive only sample.
    \deleted{The trend seen in the early-type host galaxies 
    does not seem to be present when looking at external cosmologically standardized data.
    }\added{There appears to be no systemic nor statistical bias in ages values where the \citetalias{Kang2020} regression line would go through this external cosmological data set.}
    }
    \label{fig:pantheon}
\end{figure}

For a more direct and empirical test of the \replaced{size and sign of}{average} \hr{} of \sne with young progenitors we analyzed the full sample of low redshift \sne in the Pantheon sample with host galaxy properties derived by \citet{Jones2018} ($N=254$). We measured the correlation between \hr and host age as in \citetalias{Kang2020}. \replaced{This analysis}{The Pantheon sample} used light-weighted ages derived from SED fitting via ZPEG \citep{LeBorgne2002}, as described in \citet{Jones2018}.
ZPEG uses 15 star formation histories, the Salpeter initial mass function \citep{Salpeter1955}, 200 stellar age bins, 6 metallicity bins, and marginalizes over $E(B-V)$ in order to fit the observed photometry.
\cref{fig:pantheon} shows the expected result that the majority of low redshift \sne are seen in young hosts. The \hr{s} seen in these hosts are strongly inconsistent with the $+0.25 \un{mag}$ \added{average} residual predicted by extrapolating the trend proposed by \citetalias{Kang2020}. Indeed, only a small number of all \sne (at any age) show residuals of $\gtrsim0.25 \un{mag}$, contrary to the prediction that this is the average \hr{} for \sne in young hosts.
\deleted{Since Pantheon has both young spiral and old elliptical host galaxies, it is not necessary to extrapolate outside the range of the \citetalias{Kang2020} data.
}\added{No bias in age or uncertainty (Gaussian, log-normal, or otherwise) would make the predicted trend match the data.}

\replaced{As seen in \cref{fig:pantheon}, the}{The} age trend \added{seen} in the Pantheon sample is \deleted{found to be $0.0011 \pm 0.0018 \un{mag/Gyr}$, }consistent with no trend. \deleted{at the $0.6 \sigma$ level. }Light-weighted ages are \deleted{typically }biased \replaced{younger}{young} by bright \deleted{young }stars, reducing the range of observed ages and increasing the measured slope. \added{It is difficult to quantify this bias into an uncertainty on age. As such, these should only be treated as very crude estimates.} Not surprisingly, by excluding the mass step correction the size of a trend with age more than doubles 
due to the aforementioned correlation between host mass and age, though this trend is still only  significant at the $1.4\sigma$ level, \replaced{consistent with that}{similar to what was} seen by \citet{Rose2019} for the SDSS data.  
When using the same light-curve standardization parameters ($\alpha=0.15$, $\beta=3.69$) as \citetalias{Kang2020}, but including the mass step and BBC corrections, the correlation only has a $0.6\sigma$ significance. If we restrict ourselves to \replaced{the oldest galaxies}{early-type galazies}, as is the sample in \citetalias{Kang2020}, a very weak trend is found ($1.2\sigma$\deleted{ with only early-type hosts}). 
No method of examining the Pantheon data set, was able to find a significant uncorrected trend with age.
\deleted{Our seventh conclusion is the combination of our fifth and sixth: when looking at a full cosmological data set, we see no significant trend and therefore we see no evidence for a significant and unaccounted for bias in the dark energy signal from \sn.}

\added{We conclude that the linear extrapolation to young ages is inconsistent with external data. Seeing no significant trend in a cosmological data set, we find no evidence for a significant unaccounted for bias in the cosmic acceleration signal from \sn.}

\section{Conclusions}

\citet{Kang2020} claim that an empirically-determined dependence of \sn host age and luminosity derived from a small sample of early-type host galaxies can be extrapolated to large samples and young ages to account for the majority of the \replaced{dark energy}{cosmic acceleration} signal. However, we find that this trend is not robust to reanalysis. The first issue is that $12\%$ of the final sample, would not pass the \deleted{standard }JLA cosmological quality cuts, meaning that a large fraction of the data does not have reliable \hr{s}.

The inclusion of standard error sources, clearly present in \sn residuals, reduces the significance of the dependence to $< 2 \sigma$. Bypassing any need for formal uncertainty accounting, the Spearman rank-order coefficient only sees a correlation at $2\sigma$. Further, the removal of the single \sn with the oldest host and a poorly sampled light curve, SN2003ic, reduces the significance to $1.5\sigma$. Finally, by doing a full re-fit and \added{Bayesian hierarchical} analysis that also marginalizes over the correlations in the standardization coefficients\deleted{ with a Bayesian hierarchical model}, we find the trend falls to a $1.5\sigma$ \replaced{and}{or} $0.6\sigma$ significance with and without SN2003ic respectively. \deleted{The certainty and significance of this trend appear to not be robust.}

If this correlation exists, the propagation to a bias in cosmological parameters is not direct or simple. \replaced{Stellar ages are notoriously difficult to measure, and as such extra uncertainty should be added when propagating any host galaxy age correlation across cosmic time. For example, we find that the YEPS uncertainties appear to be under estimated, on average, by $1.6 \pm 0.3 \un{Gyr}$.}{When replacing the \hr{s} from \citetalias{Kang2020} with those used in the Pantheon analysis, we see that the standard practice of applying a host galaxy mass correction leaves only a very weak and insignificant relation between \hr{s} and inferred age.}

Finally, comparing the claimed trend against large, recent cosmological samples, which include young hosts, the trend is strongly ruled out. \deleted{We see that the standard practice of applying a host galaxy mass correction leaves only a very weak and insignificant relation between \hr{s} and inferred age. Though there is evidence for a small host galaxy dependence of \sn peak luminosity, going back over 10 years, we find no evidence to question the more than 20 years of dark energy measurements.}

The recent results of \citetalias{Kang2020}, upon re-examination, do not justify calling into question the presence of dark energy. However, we do concur with their closing remarks: the redshift dependence of \sn remains an important challenge for future precision dark energy measurements and requires ongoing studies.

\acknowledgements
The authors \deleted{would like to }thank Peter Garnavich, Young-Wook Lee, and Young-Lo Kim for their time reading and commenting on an early draft.
\added{In addition, we thank the two referees whose comments improved the clarity, and hopefully impact, of this paper.}
B.M.R., D.R., A.C., S.E.D., S.D., and A.F. acknowledge support from NASA through grant NNG16PJ311I.
D.O.J. is supported by a Gordon and Betty Moore Foundation postdoctoral fellowship at the University of California, Santa Cruz.
D.M.S. is supported by DOE grant de-sc0010007, the David and Lucile Packard Foundation, and in part by NASA through grant NNG17PX03C.

\software{
corner.py \citep{Foreman-Mackey2016}, 
kde\_corner,
Matplotlib \citep{matplotlib}, 
Numpy \citep{numpy}, 
Pandas \citep{pandas}, 
pystan (\doi{10.5281/zenodo.598257}),
python, 
SciPy \citep{scipy}, 
stan \citep{Carpenter2017},
UNITY \citep{Rubin2015},
ZPEG \citep{LeBorgne2002}
}

\vspace{1em}
\bibliographystyle{apj}
\bibliography{library}

\begin{thebibliography}{}
\expandafter\ifx\csname natexlab\endcsname\relax\def\natexlab#1{#1}\fi

\bibitem[{Betoule {et~al.}(2014)Betoule, Kessler, Guy, Mosher, Hardin, Biswas,
  Astier, {El-Hage}, Konig, Kuhlmann, Marriner, Pain, Regnault, Balland,
  Bassett, Brown, Campbell, Carlberg, {Cellier-Holzem}, Cinabro, Conley,
  D'Andrea, Depoy, Doi, Ellis, Fabbro, Filippenko, Foley, Frieman, Fouchez,
  Galbany, Goobar, Gupta, Hill, Hlozek, Hogan, Hook, Howell, Jha, Le~Guillou,
  Leloudas, Lidman, Marshall, M{\"o}ller, Mour{\~a}o, Neveu, Nichol, Olmstead,
  {Palanque-Delabrouille}, Perlmutter, Prieto, Pritchet, Richmond, Riess,
  {Ruhlmann-Kleider}, Sako, Schahmaneche, Schneider, Smith, Sollerman,
  Sullivan, Walton, \& Wheeler}]{Betoule2014}
Betoule, M., Kessler, R., Guy, J., {et~al.} 2014, A\&A, 568, A22

\bibitem[{Burns {et~al.}(2011)Burns, Stritzinger, Phillips, Kattner, Persson,
  Madore, Freedman, Boldt, Campillay, Contreras, Folatelli, Gonzalez,
  Krzeminski, Morrell, Salgado, \& Suntzeff}]{Burns2011}
Burns, C.~R., Stritzinger, M., Phillips, M.~M., {et~al.} 2011, ApJ, 141, 19

\bibitem[{Campbell {et~al.}(2013)Campbell, D'Andrea, Nichol, Sako, Smith,
  Lampeitl, Olmstead, Bassett, Biswas, Brown, Cinabro, Dawson, Dilday, Foley,
  Frieman, Garnavich, Hlozek, Jha, Kuhlmann, Kunz, Marriner, Miquel, Richmond,
  Riess, Schneider, Sollerman, Taylor, \& Zhao}]{Campbell2013}
Campbell, H., D'Andrea, C.~B., Nichol, R.~C., {et~al.} 2013, ApJ, 763, 88

\bibitem[{Carpenter {et~al.}(2017)Carpenter, Gelman, Hoffman, Lee, Goodrich,
  Betancourt, Brubaker, Guo, Li, \& Riddell}]{Carpenter2017}
Carpenter, B., Gelman, A., Hoffman, M.~D., {et~al.} 2017, Journal of
  Statistical Software, 76, 1

\bibitem[{Childress {et~al.}(2014)Childress, Wolf, \& Zahid}]{Childress2014}
Childress, M.~J., Wolf, C., \& Zahid, H.~J. 2014, MNRAS, 445, 1898

\bibitem[{Chung {et~al.}(2013)Chung, Yoon, Lee, \& Lee}]{Chung2013}
Chung, C., Yoon, S.-J., Lee, S.-Y., \& Lee, Y.-W. 2013, ApJS, 204, 3

\bibitem[{Cikota {et~al.}(2019)Cikota, Patat, Wang, Wheeler, Bulla, Baade,
  H{\"o}flich, Cikota, Clocchiatti, Maund, Stevance, \& Yang}]{Cikota2019}
Cikota, A., Patat, F., Wang, L., {et~al.} 2019, MNRAS, 490, 578

\bibitem[{{DES Collaboration} {et~al.}(2019){DES Collaboration}, Abbott, Allam,
  Andersen, Angus, Asorey, Avelino, Avila, Bassett, Bechtol, Bernstein, Bertin,
  Brooks, Brout, Brown, Burke, Calcino, Rosell, Carollo, Carrasco~Kind,
  Carretero, Casas, Castander, Cawthon, Challis, Childress, Clocchiatti, Cunha,
  D'Andrea, {da Costa}, Davis, Davis, De~Vicente, DePoy, Desai, Diehl, Doel,
  {Drlica-Wagner}, Eifler, Evrard, Fernandez, Filippenko, Finley, Flaugher,
  Foley, Fosalba, Frieman, Galbany, {Garc{\'i}a-Bellido}, Gaztanaga,
  Giannantonio, Glazebrook, Goldstein, {Gonz{\'a}lez-Gait{\'a}n}, Gruen,
  Gruendl, Gschwend, Gupta, Gutierrez, Hartley, Hinton, Hollowood, Honscheid,
  Hoormann, Hoyle, James, Jeltema, Johnson, Johnson, Kasai, Kent, Kessler, Kim,
  Kirshner, Kovacs, Krause, Kron, Kuehn, Kuhlmann, Kuropatkin, Lahav, Lasker,
  Lewis, Li, Lidman, Lima, Lin, Macaulay, Maia, Mandel, March, Marriner,
  Marshall, Martini, Menanteau, Miller, Miquel, Miranda, Mohr, Morganson,
  Muthukrishna, M{\"o}ller, Neilsen, Nichol, Nord, Nugent, Ogando, Palmese,
  Pan, Plazas, Pursiainen, Romer, Roodman, Rozo, Rykoff, Sako, Sanchez,
  Scarpine, Schindler, Schubnell, Scolnic, Serrano, {Sevilla-Noarbe}, Sharp,
  Smith, {Soares-Santos}, Sobreira, Sommer, Spinka, Suchyta, Sullivan, Swann,
  Tarle, Thomas, Thomas, Troxel, Tucker, Uddin, Walker, Wester, Wiseman, Wolf,
  Yanny, Zhang, \& Zhang}]{DESCollaboration2019}
{DES Collaboration}, Abbott, T. M.~C., Allam, S., {et~al.} 2019, ApJL, 872, L30

\bibitem[{Faber {et~al.}(1992)Faber, Worthey, \& Gonzales}]{Faber1992}
Faber, S.~M., Worthey, G., \& Gonzales, J.~J. 1992, in {{IAU}} Symposium, Vol.
  149, The Stellar Populations of Galaxies, ed. B.~Barbuy \& A.~Renzini, 255

\bibitem[{Foley {et~al.}(2013)Foley, Challis, Chornock, Ganeshalingam, Li,
  Marion, Morrell, Pignata, Stritzinger, Silverman, Wang, Anderson, Filippenko,
  Freedman, Hamuy, Jha, Kirshner, McCully, Persson, Phillips, Reichart, \&
  Soderberg}]{Foley2013}
Foley, R.~J., Challis, P.~J., Chornock, R., {et~al.} 2013, ApJ, 767, 57

\bibitem[{{Foreman-Mackey}(2016)}]{Foreman-Mackey2016}
{Foreman-Mackey}, D. 2016, JOSS, 24, doi:10.21105/joss.00024

\bibitem[{Freedman {et~al.}(2019)Freedman, Madore, Hatt, Hoyt, Jang, Beaton,
  Burns, Lee, Monson, Neeley, Phillips, Rich, \& Seibert}]{Freedman2019}
Freedman, W.~L., Madore, B.~F., Hatt, D., {et~al.} 2019, ApJ, 882, 34

\bibitem[{Gallagher {et~al.}(2008)Gallagher, Garnavich, Caldwell, Kirshner,
  Jha, Li, Ganeshalingam, \& Filippenko}]{Gallagher2008}
Gallagher, J.~S., Garnavich, P.~M., Caldwell, N., {et~al.} 2008, ApJ, 685, 752

\bibitem[{Garnavich {et~al.}(1998)Garnavich, Jha, Challis, Clocchiatti,
  Diercks, Filippenko, Gilliland, Hogan, Kirshner, Leibundgut, Phillips, Reiss,
  Riess, Schmidt, Schommer, Smith, Spyromilio, Stubbs, Suntzeff, Tonry, \&
  Carroll}]{Garnavich1998b}
Garnavich, P.~M., Jha, S., Challis, P., {et~al.} 1998, ApJ, 509, 74

\bibitem[{Garnavich {et~al.}(2004)Garnavich, Bonanos, Krisciunas, Jha,
  Kirshner, Schlegel, Challis, Macri, Hatano, Branch, Bothun, \&
  Freedman}]{Garnavich2004}
Garnavich, P.~M., Bonanos, A.~Z., Krisciunas, K., {et~al.} 2004, ApJ, 613, 1120

\bibitem[{Gupta {et~al.}(2011)Gupta, D'Andrea, Sako, Conroy, Smith, Bassett,
  Frieman, Garnavich, Jha, Kessler, Lampeitl, Marriner, Nichol, \&
  Schneider}]{Gupta2011}
Gupta, R.~R., D'Andrea, C.~B., Sako, M., {et~al.} 2011, ApJ, 740, 92

\bibitem[{Guy {et~al.}(2007)Guy, Astier, Baumont, Hardin, Pain, Regnault, Basa,
  Carlberg, Conley, Fabbro, Fouchez, Hook, Howell, Perrett, Pritchet, Rich,
  Sullivan, Antilogus, Aubourg, Bazin, Bronder, Filiol,
  {Palanque-Delabrouille}, Ripoche, \& {Ruhlmann-Kleider}}]{Guy2007}
Guy, J., Astier, P., Baumont, S., {et~al.} 2007, A\&A, 466, 11

\bibitem[{Guy {et~al.}(2010)Guy, Sullivan, Conley, Regnault, Astier, Balland,
  Basa, Carlberg, Fouchez, Hardin, Hook, Howell, Pain, {Palanque-Delabrouille},
  Perrett, Pritchet, Rich, {Ruhlmann-Kleider}, Balam, Baumont, Ellis, Fabbro,
  Fakhouri, Fourmanoit, {Gonz{\'a}lez-Gait{\'a}n}, Graham, Hsiao, Kronborg,
  Lidman, Mourao, Perlmutter, Ripoche, Suzuki, \& Walker}]{Guy2010}
Guy, J., Sullivan, M., Conley, A., {et~al.} 2010, A\&A, 523, A7

\bibitem[{Hamuy {et~al.}(1995)Hamuy, Phillips, Maza, Suntzeff, Schommer, \&
  Aviles}]{Hamuy1995}
Hamuy, M., Phillips, M.~M., Maza, J., {et~al.} 1995, AJ, 109, 1

\bibitem[{Hamuy {et~al.}(2000)Hamuy, Trager, Pinto, Phillips, Schommer, Ivanov,
  \& Suntzeff}]{Hamuy2000}
Hamuy, M., Trager, S.~C., Pinto, P.~A., {et~al.} 2000, AJ, 120, 1479

\bibitem[{Hamuy {et~al.}(1996)Hamuy, Phillips, Suntzeff, Schommer, Maza,
  Antezan, Wischnjewsky, Valladares, Muena, Gonzales, Aviles, Wells, Smith,
  Navarrete, Covarrubias, Williger, Walker, Layden, Elias, Baldwin, Hernandez,
  Tirado, Ugarte, Elston, Saavedra, Barrientos, Costa, Lira, Ruiz, Anguita,
  Gomez, Ortiz, {della Valle}, Danziger, Storm, Kim, Bailyn, Rubenstein,
  Tucker, Cersosimo, Mendez, Siciliano, Sherry, Chaboyer, Koopmann, Geisler,
  Sarajedini, Dey, Tyson, Rich, Gal, Lamontagne, Caldwell, Guhathakurta,
  Phillips, Szkody, Prosser, Ho, McMahan, Baggley, Cheng, Havlen, Wakamatsu,
  Janes, Malkan, Baganoff, Seitzer, Shara, Sturch, Hesser, Hartig, Hughes,
  Welch, Williams, Ferguson, Francis, French, Bolte, Roth, Odewahn, Howell, \&
  Krzeminski}]{Hamuy1996d}
Hamuy, M., Phillips, M.~M., Suntzeff, N.~B., {et~al.} 1996, AJ, 112, 2408

\bibitem[{Hicken {et~al.}(2009)Hicken, {Wood-Vasey}, Blondin, Challis, Jha,
  Kelly, Rest, \& Kirshner}]{Hicken2009}
Hicken, M., {Wood-Vasey}, W.~M., Blondin, S., {et~al.} 2009, ApJ, 700, 1097

\bibitem[{Hunter(2007)}]{matplotlib}
Hunter, J.~D. 2007, Computing in Science \& Engineering, 9, 90

\bibitem[{Jha {et~al.}(2007)Jha, Riess, \& Kirshner}]{Jha2007}
Jha, S., Riess, A.~G., \& Kirshner, R.~P. 2007, ApJ, 659, 122

\bibitem[{Jones {et~al.}(2015)Jones, Riess, \& Scolnic}]{Jones2015}
Jones, D.~O., Riess, A.~G., \& Scolnic, D.~M. 2015, ApJ, 812, 31

\bibitem[{Jones {et~al.}(2018)Jones, Riess, Scolnic, Pan, Johnson, Coulter,
  Dettman, Foley, Foley, Huber, Jha, Kilpatrick, Kirshner, Rest, Schultz, \&
  Siebert}]{Jones2018}
Jones, D.~O., Riess, A.~G., Scolnic, D.~M., {et~al.} 2018, ApJ, 867, 108

\bibitem[{Jones {et~al.}(2019)Jones, Scolnic, Foley, Rest, Kessler, Challis,
  Chambers, Coulter, Dettman, Foley, Huber, Jha, Johnson, Kilpatrick, Kirshner,
  Manuel, Narayan, Pan, Riess, Schultz, Siebert, Berger, Chornock, Flewelling,
  Magnier, Smartt, Smith, Wainscoat, Waters, \& Willman}]{Jones2019}
Jones, D.~O., Scolnic, D.~M., Foley, R.~J., {et~al.} 2019, ApJ, 881, 19

\bibitem[{Jones {et~al.}(2001)Jones, Oliphant, Peterson, {et~al.}}]{scipy}
Jones, E., Oliphant, T., Peterson, P., {et~al.} 2001, arXiv:1907.10121

\bibitem[{Kang {et~al.}(2016)Kang, Kim, Lim, Chung, \& Lee}]{Kang2016}
Kang, Y., Kim, Y.-L., Lim, D., Chung, C., \& Lee, Y.-W. 2016, ApJS, 223

\bibitem[{Kang {et~al.}(2020)Kang, Lee, Kim, Chung, \& Ree}]{Kang2020}
Kang, Y., Lee, Y.-W., Kim, Y.-L., Chung, C., \& Ree, C.~H. 2020, ApJ, 889, 8

\bibitem[{Kelly(2007)}]{Kelly2007a}
Kelly, B.~C. 2007, 665, 1489

\bibitem[{Kelly {et~al.}(2010)Kelly, Hicken, Burke, Mandel, \&
  Kirshner}]{Kelly2010}
Kelly, P.~L., Hicken, M., Burke, D.~L., Mandel, K.~S., \& Kirshner, R.~P. 2010,
  ApJ, 715, 743

\bibitem[{Kessler \& Scolnic(2017)}]{Kessler2017}
Kessler, R., \& Scolnic, D. 2017, ApJ, 836, 56

\bibitem[{Kim {et~al.}(2019)Kim, Kang, \& Lee}]{Kim2019}
Kim, Y.-L., Kang, Y., \& Lee, Y.-W. 2019, Journal of Korean Astronomical
  Society, 52, 181

\bibitem[{Kim {et~al.}(2018)Kim, Smith, Sullivan, \& Lee}]{Kim2018}
Kim, Y.-L., Smith, M., Sullivan, M., \& Lee, Y.-W. 2018, ApJ, 854, 24

\bibitem[{Kowal(1968)}]{Kowal1968}
Kowal, C.~T. 1968, AJ, 73, 1021

\bibitem[{Kunz {et~al.}(2007)Kunz, Bassett, \& Hlozek}]{Kunz2007}
Kunz, M., Bassett, B.~A., \& Hlozek, R.~A. 2007, Phys. Rev. D, 75, 103508

\bibitem[{Lampeitl {et~al.}(2010)Lampeitl, Smith, Nichol, Bassett, Cinabro,
  Dilday, Foley, Frieman, Garnavich, Goobar, Im, Jha, Marriner, Miquel, Nordin,
  Ostman, Riess, Sako, Schneider, Sollerman, \& Stritzinger}]{Lampeitl2010}
Lampeitl, H., Smith, M., Nichol, R.~C., {et~al.} 2010, ApJ, 722, 566

\bibitem[{Le~Borgne \& {Rocca-Volmerange}(2002)}]{LeBorgne2002}
Le~Borgne, D., \& {Rocca-Volmerange}, B. 2002, A\&A, 386, 446

\bibitem[{McKinney(2010)}]{pandas}
McKinney, W. 2010, Data {{Structures}} for {{Statistical Computing}} in
  {{Python}}

\bibitem[{{Moreno-Raya} {et~al.}(2016){Moreno-Raya}, {L{\'o}pez-S{\'a}nchez},
  Moll{\'a}, Galbany, V{\'i}lchez, \& Carnero}]{Moreno-Raya2016a}
{Moreno-Raya}, M.~E., {L{\'o}pez-S{\'a}nchez}, {\'A}.~R., Moll{\'a}, M.,
  {et~al.} 2016, MNRAS, 462, 1281

\bibitem[{Mosher {et~al.}(2014)Mosher, Guy, Kessler, Astier, Marriner, Betoule,
  Sako, {El-Hage}, Biswas, Pain, Kuhlmann, Regnault, Frieman, \&
  Schneider}]{Mosher2014}
Mosher, J., Guy, J., Kessler, R., {et~al.} 2014, ApJ, 793, 16

\bibitem[{Perlmutter {et~al.}(1997)Perlmutter, Gabi, Goldhaber, Goobar, Groom,
  Hook, Kim, Kim, Lee, Pain, Pennypacker, Small, Ellis, McMahon, Boyle,
  Bunclark, Carter, Irwin, Glazebrook, Newberg, Filippenko, Matheson, Dopita,
  \& Couch}]{Perlmutter1997}
Perlmutter, S., Gabi, S., Goldhaber, G., {et~al.} 1997, ApJ, 483, 565

\bibitem[{Perlmutter {et~al.}(1999)Perlmutter, Aldering, Goldhaber, Knop,
  Nugent, Castro, Deustua, Fabbro, Goobar, Groom, Hook, Kim, Kim, Lee, Nunes,
  Pain, Pennypacker, Quimby, Lidman, Ellis, Irwin, McMahon, Ruiz-Lapuente,
  Walton, Schaefer, Boyle, Filippenko, Matheson, Fruchter, Panagia, Newberg,
  Couch, \& Project}]{Perlmutter1999}
Perlmutter, S., Aldering, G., Goldhaber, G., {et~al.} 1999, ApJ, 517, 565

\bibitem[{Phillips(1993)}]{Phillips1993}
Phillips, M.~M. 1993, ApJL, 413, L105

\bibitem[{Pskovskii(1969)}]{Pskovskii1969}
Pskovskii, Y.~P. 1969, Soviet Astronomy, 12, 750

\bibitem[{Pskovskii(1977)}]{Pskovskii1977}
---. 1977, SvA, 21, 675

\bibitem[{Rest {et~al.}(2014)Rest, Scolnic, Foley, Huber, Chornock, Narayan,
  Tonry, Berger, Soderberg, Stubbs, Riess, Kirshner, Smartt, Schlafly, Rodney,
  Botticella, Brout, Challis, Czekala, Drout, Hudson, Kotak, Leibler, Lunnan,
  Marion, McCrum, Milisavljevic, Pastorello, Sanders, Smith, Stafford, Thilker,
  Valenti, {Wood-Vasey}, Zheng, Burgett, Chambers, Denneau, Draper, Flewelling,
  Hodapp, Kaiser, Kudritzki, Magnier, Metcalfe, Price, Sweeney, Wainscoat, \&
  Waters}]{Rest2014}
Rest, A., Scolnic, D., Foley, R.~J., {et~al.} 2014, ApJ, 795, 44

\bibitem[{Riess {et~al.}(1996)Riess, Press, \& Kirshner}]{Riess1996}
Riess, A.~G., Press, W.~H., \& Kirshner, R.~P. 1996, ApJ, 473, 88

\bibitem[{Riess {et~al.}(1998)Riess, Filippenko, Challis, Clocchiattia,
  Diercks, Garnavich, Gilliland, Hogan, Jha, Kirshner, Leibundgut, Phillips,
  Reiss, Schmidt, Schommer, Smith, Spyromilio, Stubbs, Suntzeff, \&
  Tonry}]{Riess1998}
Riess, A.~G., Filippenko, A.~V., Challis, P., {et~al.} 1998, ApJ, 116, 1009

\bibitem[{Riess {et~al.}(2018)Riess, Casertano, Yuan, Macri, Bucciarelli,
  Lattanzi, MacKenty, Bowers, Zheng, Filippenko, Huang, \&
  Anderson}]{Riess2018b}
Riess, A.~G., Casertano, S., Yuan, W., {et~al.} 2018, ApJ, 861, 126

\bibitem[{Rigault {et~al.}(2013)Rigault, Copin, Aldering, Antilogus, Aragon,
  Bailey, Baltay, Bongard, Buton, Canto, {Cellier-Holzem}, Childress, Chotard,
  Fakhouri, Feindt, Fleury, Gangler, Greskovic, Guy, Kim, Kowalski, Lombardo,
  Nordin, Nugent, Pain, Pecontal, Pereira, Perlmutter, Rabinowitz, Runge,
  Saunders, Scalzo, Smadja, Tao, Thomas, Weaver, \& Factory)}]{Rigault2013}
Rigault, M., Copin, Y., Aldering, G., {et~al.} 2013, A\&A, 560, A66

\bibitem[{Rigault {et~al.}(2018)Rigault, Brinnel, Aldering, Antilogus, Aragon,
  Bailey, Baltay, Barbary, Bongard, Boone, Buton, Childress, Chotard, Copin,
  Dixon, Fagrelius, Feindt, Fouchez, Gangler, Hayden, Hillebrandt, Howell, Kim,
  Kowalski, Kuesters, Leget, Lombardo, Lin, Nordin, Pain, Pecontal, Pereira,
  Perlmutter, Rabinowitz, Runge, Rubin, Saunders, Smadja, Sofiatti, Suzuki,
  Taubenberger, Tao, \& Thomas}]{Rigault2018}
Rigault, M., Brinnel, V., Aldering, G., {et~al.} 2018, arXiv:1806.03849

\bibitem[{Rose {et~al.}(2019)Rose, Garnavich, \& Berg}]{Rose2019}
Rose, B.~M., Garnavich, P.~M., \& Berg, M.~A. 2019, ApJ, 874, 32

\bibitem[{Rose {et~al.}(2020)Rose, Dixon, Rubin, Hounsell, Saunders, Deustua,
  Fruchter, Galbany, Perlmutter, \& Sako}]{Rose2020a}
Rose, B.~M., Dixon, S., Rubin, D., {et~al.} 2020, ApJ, 890, 60

\bibitem[{Rubin {et~al.}(2015)Rubin, Aldering, Barbary, Boone, Chappell,
  Currie, Deustua, Fagrelius, Fruchter, Hayden, Lidman, Nordin, Perlmutter,
  Saunders, \& Sofiatti}]{Rubin2015}
Rubin, D., Aldering, G., Barbary, K., {et~al.} 2015, ApJ, 813, 137

\bibitem[{Rust(1974)}]{Rust1974}
Rust, B.~W. 1974, PhD thesis, Oak Ridge National Lab., TN.

\bibitem[{Sako {et~al.}(2008)Sako, Bassett, Becker, Cinabro, DeJongh, Depoy,
  Dilday, Doi, Frieman, Garnavich, Hogan, Holtzman, Jha, Kessler, Konishi,
  Lampeitl, Marriner, Miknaitis, Nichol, Prieto, Riess, Richmond, Romani,
  Schneider, Smith, SubbaRao, Takanashi, Tokita, van~der Heyden, Yasuda, Zheng,
  Barentine, Brewington, Choi, Dembicky, Harnavek, Ihara, Im, Ketzeback,
  Kleinman, Krzesi{\'n}ski, Long, Malanushenko, Malanushenko, McMillan,
  Morokuma, Nitta, Pan, Saurage, \& Snedden}]{Sako2008}
Sako, M., Bassett, B., Becker, A., {et~al.} 2008, AJ, 135, 348

\bibitem[{Sako {et~al.}(2018)Sako, Bassett, Becker, Brown, Campbell, Cane,
  Cinabro, D'Andrea, Dawson, DeJongh, Depoy, Dilday, Doi, Filippenko, Fischer,
  Foley, Frieman, Galbany, Garnavich, Goobar, Gupta, Hill, Hayden, Hlozek,
  Holtzman, Hopp, Jha, Kessler, Kollatschny, Leloudas, Marriner, Marshall,
  Miquel, Morokuma, Mosher, Nichol, Nordin, Olmstead, Ostman, Prieto, Richmond,
  Romani, Sollerman, Stritzinger, Schneider, Smith, Wheeler, Yasuda, \&
  Zheng}]{Sako2014}
Sako, M., Bassett, B., Becker, A.~C., {et~al.} 2018, PASP, 130, 064002

\bibitem[{Salpeter(1955)}]{Salpeter1955}
Salpeter, E.~E. 1955, ApJ, 121, 161

\bibitem[{Scolnic \& Kessler(2016)}]{Scolnic2016}
Scolnic, D., \& Kessler, R. 2016, ApJL, 822, L35

\bibitem[{Scolnic {et~al.}(2018)Scolnic, Jones, Rest, Pan, Chornock, Foley,
  Huber, Kessler, Narayan, Riess, Rodney, Berger, Challis, Drout, Finkbeiner,
  Lunnan, Kirshner, Sanders, Schlafly, Smartt, Stubbs, Tonry, {Wood-Vasey},
  Foley, Hand, Johnson, Burgett, Chambers, Draper, Hodapp, Kaiser, Kudritzki,
  Magnier, Metcalfe, Bresolin, Kotak, McCrum, \& Smith}]{Scolnic2018}
Scolnic, D.~M., Jones, D.~O., Rest, A., {et~al.} 2018, ApJ, 859, 101

\bibitem[{Smith {et~al.}(2020)Smith, Sullivan, Wiseman, Kessler, Scolnic,
  Brout, D'Andrea, Davis, Foley, Frohmaier, Galbany, Gupta, Guti{\'e}rrez,
  Hinton, Kelsey, Lidman, Macaulay, M{\"o}ller, Nichol, Nugent, Palmese,
  Pursiainen, Sako, Thomas, Tucker, Carollo, Lewis, Sommer, Abbott, Aguena,
  Allam, Avila, Bertin, Bhargava, Brooks, {Buckley-Geer}, Burke, Rosell, Kind,
  Costanzi, {da Costa}, De~Vicente, Desai, Diehl, Doel, Eifler, Everett,
  Flaugher, Fosalba, Frieman, {Garc{\'i}a-Bellido}, Gaztanaga, Glazebrook,
  Gruen, Gruendl, Gschwend, Gutierrez, Hartley, Hollowood, Honscheid, James,
  Krause, Kuehn, Kuropatkin, Lima, MacCrann, Maia, Marshall, Martini, Melchior,
  Menanteau, Miquel, {Paz-Chinch{\'o}n}, Plazas, Romer, Roodman, Rykoff,
  Sanchez, Scarpine, Schubnell, Serrano, {Sevilla-Noarbe}, Suchyta, Swanson,
  Tarle, Thomas, Tucker, Varga, Walker, \& Collaboration}]{Smith2020}
Smith, M., Sullivan, M., Wiseman, P., {et~al.} 2020, arXiv:2001.11294

\bibitem[{Sullivan {et~al.}(2010)Sullivan, Conley, Howell, Neill, Astier,
  Balland, Basa, Carlberg, Fouchez, Guy, Hardin, Hook, Pain,
  {Palanque-Delabrouille}, Perrett, Pritchet, Regnault, Rich,
  {Ruhlmann-Kleider}, Baumont, Hsiao, Kronborg, Lidman, Perlmutter, \&
  Walker}]{Sullivan2010}
Sullivan, M., Conley, A., Howell, D.~A., {et~al.} 2010, MNRAS, 406, 782

\bibitem[{Suzuki {et~al.}(2012)Suzuki, Rubin, Lidman, Aldering, Amanullah,
  Barbary, Barrientos, Botyanszki, Brodwin, Connolly, Dawson, Dey, Doi,
  Donahue, Deustua, Eisenhardt, Ellingson, Faccioli, Fadeyev, Fakhouri,
  Fruchter, Gilbank, Gladders, Goldhaber, Gonzalez, Goobar, Gude, Hattori,
  Hoekstra, Hsiao, Huang, Ihara, Jee, Johnston, Kashikawa, Koester, Konishi,
  Kowalski, Linder, Lubin, Melbourne, Meyers, Morokuma, Munshi, Mullis, Oda,
  Panagia, Perlmutter, Postman, Pritchard, Rhodes, Ripoche, Rosati, Schlegel,
  Spadafora, Stanford, Stanishev, Stern, Strovink, Takanashi, Tokita, Wagner,
  Wang, Yasuda, \& Yee}]{Suzuki2012}
Suzuki, N., Rubin, D., Lidman, C., {et~al.} 2012, ApJ, 746, 85

\bibitem[{Tripp(1998)}]{Tripp1998}
Tripp, R. 1998, A\&A, 331, 815

\bibitem[{Uddin {et~al.}(2017)Uddin, Mould, Lidman, {Ruhlmann-Kleider}, \&
  Zhang}]{Uddin2017a}
Uddin, S.~A., Mould, J., Lidman, C., {Ruhlmann-Kleider}, V., \& Zhang, B.~R.
  2017, ApJ, 848, 56

\bibitem[{{van der Walt} {et~al.}(2011){van der Walt}, Colbert, \&
  Varoquaux}]{numpy}
{van der Walt}, S., Colbert, S.~C., \& Varoquaux, G. 2011, CSE, 13, 22

\bibitem[{Worthey {et~al.}(1994)Worthey, Faber, Gonzalez, \&
  Burstein}]{Worthey1994}
Worthey, G., Faber, S.~M., Gonzalez, J.~J., \& Burstein, D. 1994, 94, 687

\end{thebibliography}

\listofchanges

\end{document}